\documentstyle[12pt,psfig]{article}

\begin{document}
\begin{center}
\LARGE{A Strange Star Model for Pulsars\\

R. X. Xu$^{1,2}$, G. J. Qiao$^{1,2}$}

\end{center}

{$^1$Department of Geophysics, Peking University, 
 Beijing 100871, China\\
$^2$Chinese Academy of Sciences-Peking University joint Beijing\\
 Astrophysical Center\\

Email: rxxu@bac.pku.edu.cn, gjn@pku.edu.cn}\\
{\it Sent to Astrophysics Journal (part 1) on April 11, 1998}
\date{1998.4.11}
\newpage

\begin{abstract}

It is suggested in this paper that the `bare' strange star might be not bare, and there could be a magnetosphere around it. As a strange star might be an intensely magnetized rotator, the induced unipolar electric field would be large enough to construct a magnetosphere around the strange matter core. This kind of magnetosphere is very similar to that of the rotating magnetized neutron stars discussed by many authors. A magnetosphere will be established very soon through pair production by $\gamma-B$ or two photon processes after a strange star was born in a supernova explosion. It is emphasized that the fact that the strange star surface can not supply charged particles does not stop the formation of a space charge separated magnetosphere around the bare strange star. An accretion crust is quite difficult to come into being around an isolated strange star. Therefore the observed radio signals of an rotation-powered pulsar may come from a bare strange stars rather than a neutron stars or a strange star with an accretion crust. The idea, that the radio pulsars are the strange stars without crusts, is supported by some observations. For example, the electron-positron annihilation line in the spectrum of the Crab pulsar has been reported (Agrinier et al. 1990; Massaro et al. 1991); the iron emission lines have been observed in many X-ray pulsars but never been reported in $X-$ray emission of radio pulsars. This fact is difficult to be understood if the radio pulsars are the neutron stars where the surface binding energy of iron ions is too low to avoid a ion free-flow from the surface (Neuhauseret et al.1986, 1987).

\end{abstract}

elementary particles - pulsars: general - stars: neutron - hydromagnetics

\section{Introduction}

Since the first radio pulsar, CP 1919, was discovered in November 1967 (Hewish, Bell, et al. 1968), more and more radio pulsars are found, the number of which reaches about 750 (Becker \& Trumper 1997). These objects are now almost universally believed to be rotating magnetized neutron stars. However, it was suggested that there might be no neutron star, only strange stars (e.g. Alock et al.1986). Hence, one of the very interesting and most fundamental questions is `What is the nature of pulsars?'(i.e. Do the signals of pulsars come from neutron stars or from strange stars?). Unfortunately, the question could still not be answered with certainty even now (Broderick 1998).

Soon after the discovery of pulsars, by removing the possibilities of the pulse periods due to white dwarf pulsation or duo to rapid orbital rotation (see, e.g. a review by Smith 1977), many people widely accept the concept that the pulsars are the neutron stars, which were conceived as theoretically possible stable structures in astrophysics (Landau 1932, Oppenheimer \& Volkoff 1939). Following this, many authors discussed the inner structure of neutron stars, especially the properties of possible quark phase in the neutron star core (e.g. Wang \& Lu 1984).

As the hypothesis that strange matter may be the absolute ground state of the strong interaction confined state has been raised (Bodner 1971; Witten 1984), Farhi \& Jaffe (1984) point out that the energy of strange quark matter is lower than that of matter composed by nucleus for quantum chromodynamical parameters within rather wide range. Hence, strange stars, that might be considered as the ground state of neutron stars, could exist, and the observed pulsars might be strange stars (Alcock et al. 1986, Kettner et al. 1995). Therefore, the question about the nature of pulsars, which seems to have been answered, rises again. Now, there are two kinds of candidates for the pulsar's correspondence object: one is the classical neutron star, and another is the strange star which was proposed about a decade ago.

The key point on this question is to find the difference of the behaviors of strange stars and neutron stars, both observationally and theoretically. As both the strange-star models and neutron-star models for pulsars predict the observed pulsars' mass ($\sim 1.4 M_\odot$, from double-pulsars systems) and radius ($\sim 10^6$ cm, from X-ray bursters), it is hard to differentiate these two type models in observations. The dynamically damping effects (Wang \& Lu 1984; Dai \& Lu 1996), the minimum rotation periods (Friedman \& Olinto 1989), the cooling curves (Benvenuto \& Vucetich 1991) and the vibratory mode (Broderick et al. 1998) have been discussed in detail in the literature. However, no direct observational clue has yet shown that the pulsars are neutron stars or strange stars.

While, in the terrestrial physics, to search the new state of strong interaction matter, the so-called quark-gluon plasma (QGP), is the primary goal of relativistic heavy-ion laboratory (McLerran 1986, Muller 1995). Many proposed QGP signatures have been put forward in theory and analyzed in experimental data, but the conclusion about the discovery of QGP are ambiguities. More likely, it is suggested in theory that there is a possibility of existing strange hardron cluster (Schaffner et al. 1993) and strangelet (Benvenuto \& Lugones 1995), however, no experiment has affirmed or disaffirmed the suggestion. This laboratory physics researches should be inspired if the pulsars are distinguished as strange stars rather than neutron stars. Also, the rudimental strangelet in the early universe might have implications of fundamental importance for cosmology (e.g. the dark matters, Witten 1984).

Almost all of the proposed strange star models for pulsars have addressed the case generally contemplated by most authors that the strange star core is surrounded by a normal matter crust (Alcock et. al. 1986; Kettner et al. 1995). The essential features of this core-crust structure is that the normal hadron crust with $\sim 10^{-5}M_\odot$ and the strange quark matter core with mass of $\sim 1.4 M_\odot$ and radius of $\sim 10^6$ cm are divided by a $\sim 200$ fm electric gap. It is believed that a crust can be formed during a supernova explosion (Alcock et al.1986) by accretion. However, as discussed below, after the strange star was born, a magnetosphere could be established soon, and the radiation from the open field lines region would prevent the accretion. Therefore, a crust is difficult to form beyond a newborn strange star.

It is accepted that a strange star without crust will not supply charge particles to develop a rotating space charge separated magnetosphere (Alcock et al. 1986). The reason for this is that the maximum electric field induced by a rotating magnetized dipole, $\sim 10^{11}$ V cm$^{-1}$, is negligible when being compared with the electric field at the strange matter surface, $\sim 10^{17}$ V cm$^{-1}$. Pulsar emission mechanisms, which depend on the stellar surface as a source of plasma, will not work if there is a bare quark surface (Alcock et al. 1986). Hence the bare strange stars will not be the observed pulsars. However, there are two points pointed out here:
\begin{itemize}
\item The electric field due to electron distribution near the surface decreases quickly outward, from $\sim 10^{17}$ V cm$^{-1}$ at the surface to $\sim 10^{10}$ V cm$^{-1}$ at a very small height of 10$^{-7}$ cm above the surface. Therefore the induced electric field will control most area of the magnetosphere. 
\item The magnetosphere can be established through pair production in $\gamma-B$ or two photon processes.
\end{itemize}
So, it is proposed that the bare strange stars could act as the observed radio pulsars.

Five conclusions are obtained in this paper: 1. The bare strange stars may be not bare in fact; a magnetosphere would be settled around the strange stars if the pair production process were taken into account. 2. The magnetosphere of a strange star is very similar to that of a neutron star; the radio pulsars might be the `bare' strange stars rather than the neutron stars if the strange matter hypothesis is correct. 3.Both pulsars with parallel and anti-parallel magnetic axes relative to rotational axes can be observed. 4. The idea, that the radio pulsars are the strange stars, is supported by some observations. 5. It is suspected that the strange stars with normal matter crusts are formed in binary systems; and strange stars with crusts would act as X-ray pulsars or X-ray bursters.

The structure of this paper is as follows. In section 2 the formation of the magnetosphere of the bare strange star is discussed, including the pair production processes, which is very important for the exist of magnetosphere. A comparison of properties between magnetospheres of the trange stars and of the neutron stars is presented in section 3. In section 4, the emission of strange star with a magnetosphere is discussed.
Conclusion and discussion are shown in the section 5.

\section{Formation of the magnetosphere of a bare strange star}

Why the bare strange stars are bare out? The main reason for this is that the bare surface will not supply charged particles to form a rotating space charge separated magnetosphere (Alcock 1986). However, it is suggested as follows that the magnetosphere can be formed if the pair production process is to be taken into account. 

\subsection{The induced electric field by a rotating magnetized dipole}

If the strange stars can be the candidates of pulsars, it should be strongly magnetized and rapidly rotating. The directed radiation pencil and the cyclotron absorption lines in pulsar observations show that there might be $10^{12}$ gauss magnetic field in the pulsar polar caps.

For a rotating magnetized strange star, the unipolar induction effect should be included. Maxswell equations in the frame, which corotates with the star, are (Fawley et al. 1977)
$$
\begin{array}{lll}
\bigtriangledown \cdot {\bf E} & = & 4 \pi (\eta-\eta_R),\\
\bigtriangledown \cdot {\bf B} & = & 0,\\
\bigtriangledown \times {\bf E} & = & -{1\over c}\cdot {\partial {\bf B}\over \partial t},\\
\bigtriangledown \times {\bf B} & = & {4\pi\over c}({\bf J}-{\bf J}_R) + {1\over c} \cdot {\partial {\bf E}\over \partial t},\\
\end{array}
$$
where $\eta$ is the space-charge density, ${\bf J}$ is the current density, ${\bf J}_R$ is a complicated combination of the fields and their derivatives (see the Appendix A in Fawley et al. 1977), and
$$
\eta_R = -{1\over 2\pi c}{\bf \Omega}\cdot{\bf B} + {1\over 4\pi c}({\bf \Omega}\times{\bf r})\cdot\bigtriangledown\times{\bf B}
$$
where ${\bf \Omega}$ is the angular velocity of the rotating star. If we treat a time independent problem, and simply let ${\bf J} = 0$ and ${\bf J}_R = \eta_R ({\bf \Omega}\times{\bf r})$, we come to the space charge separated density (Goldreich \& Julian 1969)
$$
\rho_{GJ} \equiv \eta_R = -{1\over 2\pi c}{\bf \Omega}\cdot{\bf B} [1-({|{\bf \Omega}\times{\bf r}| \over c})^2]^{-1},	\eqno(1)
$$
which is required for the electric field in the inertial frame to be entirely given by the corotation electric field. In the vicinity of a strange star, $|{\bf \Omega}\times{\bf r}| \ll c$, near the cap,
$$
\rho_{GJ} \sim -{{\bf \Omega}\times{\bf B}\over 2\pi c} = {1\over 9}\times 10^{-7} B_{12} P^{-1}\;\; {\rm Coulom}\; {\rm cm^{-3}}, \eqno(2)
$$
where $B_{12} = B_p/(10^{12} {\rm gauss})$ ($B_p$ is the polar cap magnetic field), and $P$ in unit of second. In Fig.4 (right), near $z = 2\times 10^{-3}$ cm, the distributed electron (bounded to the strange quark matter) charge density is comparable to the induced charge separated density, hence $\rho_{GJ}$ should be dominant when $z > 2\times 10^{-3}$ cm if there is an electric force equilibrium. As $\rho_{GJ} (\sim 10^{-8})$ is very small compared with the quark charge density $\rho_q$ ($\sim 10^{15}$, see the Appendix) in the strange star interior, it is a good approximation to think the quarks and electrons are in chemical and thermal equilibrium although charged particles have been slightly separated to balance the unipolar induced electric force in the star interior.

We can discuss the electrons distribution in the corotation frame by Thomas-Fermi model. In the star and near the surface, $\rho_{GJ}(<<\rho_q)$ could be negligible, and the boundary problem could be approximated in one dimension as [see equ.(A5)]
$$
{dV\over dz} = -\sqrt{2\alpha\over 3\pi}(V^2-\eta_R),\;\; {\rm Boundary}: V(z=0) = {3\over 4}V_q,
$$
and the equilibrium state ($z\rightarrow +\infty: {dV\over dz}\rightarrow 0$, and $V\rightarrow 0$) is not the solution of the above problem when the charge separated density $\eta_R$ (in nature unit) $\not= 0$. Hence, the electrons could not in equilibrium dynamically in the corotation frame, and also in the observers' frame.

If we have vacuum outside the strange star, the induced electric field along the magnetic field, ${\bf E}\cdot{\bf n}_B$, would be given by (Goldreich \& Julian 1969)
$$
\begin{array}{lll}
{\bf E}\cdot{\bf n}_B & = & -{2\Omega R \over c}B_p \cdot ({R\over r})^4 {\cos^3\theta \over \sqrt{1+3\cos^2\theta}}\\
& = & -1.26\times 10^{11}R_6 B_{12} P^{-1} \cdot ({R\over r})^4 {\cos^3\theta \over \sqrt{1+3\cos^2\theta}}\;\; {\rm Volt \; cm}^{-1},
\end{array} \eqno(3)
$$
where a dipole magnetic field is assumed, $B = {1\over 2}B_p \sqrt{1+3\cos^2\theta} ({R\over r})^3$, $r$ and $\theta$ are the usual polar coordinates with $\theta$ measured from the rotation axis, $R$ is the strange star radius, $R_6 = R/(10^6 {\rm cm})$, ${\bf n}_B$ is the direction of magnetic field.

In Fig.4 (left), near $z = 6\times 10^{-8}$ cm, the electric field to bound the electrons to the quark matter, $dV\over dz$, is comparable to the unipolar induced electric field along ${\bf n}_B$, hence, when $z > 6 \times 10^{-8}$ cm, the motion and distribution of electron should be mainly controlled by ${\bf E}\cdot{\bf n}_B$, as all of the other forces (e.g. gravitation and centrifugal acceleration) can be negligible (Goldreich \& Julian 1969). Thus, the distributed electrons near and above $z \sim 6\times 10^{-8}$ cm could not be mechanically or quantum mechanically equilibrium (detailed discussion below), and a magnetosphere around strange star could be established. From equ. (3) and equ.(A4), the critical height where the two electric fields are equal, ${\bf E}\cdot {\bf n}_B = {dV\over dz}$, can be obtained as a function of $\theta$, and the solution to the electric equilibrium height with $V_q = 20$ MeV is shown in Fig.1. Almost at all of the latitude degree the induced electric force can exceed that caused by strange quark matter attraction.

\begin{figure}
$$\psfig{figure=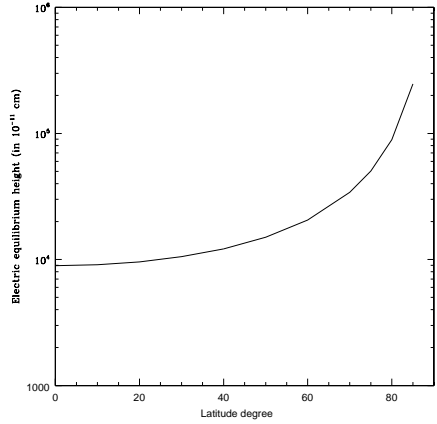,width=6cm,angle=0}$$
\caption[]{
The electric equilibrium height curve as a function of latitude degree $\theta$.
\label{fig1}}
\end{figure}

\subsection{The formation of a strange star magnetosphere}

As discussed above, the unipolar induced electric field can have considerable contribution to the distribution of electrons almost at the strange star surface ($10^{-8}$ or $10^{-3}$ cm is a very small number in the astronomical view point), which can pull or push the out part electrons near the strange star surface. The potential difference between $\theta = 0^o$ and $\theta = 90^o$ is given by (Goldreich 1972)
$$
\triangle\phi = 3 \times 10^{16} B_{12}R_6^2 P^{-1} \;\; {\rm volts},
$$
if there are no charged particles outside the star. An electron in this electric field can not be continually accelerated, as the pair creation processes could stop the acceleration. Electrons or positrons with large Lorentz factors should produce $\gamma$ rays by, for example, inverse Compton scattering, curvature radiation, synchrotron radiation and, perhaps, pair annihilation. Also most of the produced high energy $\gamma$ rays in such strong magnetic field could convert to electron-positron pairs through $\gamma + B \rightarrow e^{\pm} + B$ or two photon processes, such as $\gamma + X \rightarrow e^{\pm}$. Hence, if a strange star have a vacuum outside, this cascade of pair creation should bring about the appearance of a large enough pair plasma to construct a charge separated magnetosphere around the strange star, both a corotation part and a open field lines part (Michel 1991). According to energy conservation law, the energy of the above cascade process is from the strange star's rotation. As long as the magnetosphere has been established, the detailed discussed pulsar inner as well as outer accelerators, such as the polar gap model (Ruderman \& Sutherland 1975), the slot gap model (Arons 1983), and the out gap model (Cheng Ho \& Ruderman 1984), would work for the radio as well as higher energy photons' emission, because of the electromotive force caused by the potential difference between the center and the edge of the polar cap region.

Let's come to some details. For ${\bf \Omega}\cdot{\bf B} > 0$, the induced electric field would pull the electrons out, accelerates them ultra-relativistically. As electrons lost from the quark matter, the strange star could be positively charged, the global electric circuit (Shibata 1991) might be set up. However, this global circuit could be in quasi-equilibrium, and a small vacuum gap similar to that of RS model (Ruderman \& Sutherland 1975) could be possible near the polar cap. For ${\bf \Omega}\cdot{\bf B} < 0$, the induced electric field would push the electrons inward, and a large vacuum region should be above the polar cap. Some physical processes, such as cosmic $\gamma$ rays interaction with strong magnetic fields, or electrons (scattered by neutrinos from strange star) synchrotron radiation (jump between two Landau levels), could trigger the pair creation cascade.

Hence, a space charged limited flow (Sturock 1969) would take place, the outward accelerated particles (electrons or positions) might coherently radiate radio waves (Melrose 1995) and incoherently emit high energy photons, and the inward accelerated particles could interact with electrons and quarks electrically, which may result in the observed hot spot (Wang \& Halpern 1997). Therefore, a bare strange star could act as a radio pulsar or a $\gamma$ ray pulsar.

If a strange star forms soon after a supernova, a magnetosphere composed by ions would not be possible since a lot of very energetic outward particles and photons are near the star. However, the time scale T to form an $e^\pm$ pair plasma magnetosphere is very small. The total number of $e^\pm$ pairs in the magnetosphere might be estimated as $N_{GJ}$,
$$
N_{GJ}\sim R\int_R^{R_c} {\Omega B_p \over 2 \pi c} ({R\over r})^3 rdr \approx 7\times 10^{28} R_6^3 B_{12} P^{-1},
$$
where the radius of light cylinder $R_c = {cP\over 2\pi} \gg R$ (the radius of strange star). The mean free path $l_p$ of a photon with energy greater than $\sim 1$ MeV moving through a region of magnetic B can be estimated as (Erber 1966)
$$
l_p \sim 8.8\times 10^3 e^{3\over 4\chi}/(B \sin\alpha)\;\; {\rm cm},
$$
where $\chi$ could be approximated as 1/15 (Ruderman \& Sutherland 1975), $\alpha$ is the angle between the direction of propagation photon and the magnetic field. $l_p \sim 68$ cm for $\sin\theta \sim {1\over \gamma} \sim 10^{-5}$ ($\gamma$ is the Lorentz factor of electron). Also, the mean free length $l_e$ of electron to produce photon by curvature radiation etc. could be assumed in order of $l_p$. For the cascade processes discussed, the time scale $T$ to build up a magnetosphere might be
$$
T = ({l_p+l_e\over c}){{\rm ln}N_{GJ}\over {\rm ln}2},
$$
which is in order of $10^{-7}$ seconds for typical pulsar parameters. Considering the photon escaping and the global magnetic field structure, this time scale should not change very much. Hence, ions would have less possibility in the strange radio pulsar's magnetosphere that might be mainly consist of $e^\pm$ pairs quickly, especially in the open field lines region where large outward pressure of $e^\pm$ pairs and electromagnetic waves exists. 

While strange stars are in binaries, as the accretion pressure of wind or matter could be greater than the outward pressure from the polar cap, the accretion process should be involved, and a strange star could be accretion powered $X-$ray source. For an accreting strange star, there could be two envelope crusts shielding the two polar capes. As strange matter does not react with ions because of the Coulomb barrier (the height of which is ${3\over 4}V_q \sim 15$ MeV), there could be an electrostatic gap of thickness hundreds Fermi above the surface (Alcock et al. 1986). If the magnetic field is very strong ($B_p \sim 10^{12}$ gauss), and in case of high accretion rate (High Massive X-ray Binaries), those accreted crusts should be small, where some violent processes, such as the huge release of gravitational energy and the thermal nuclear reactions, could take place. In this case, the ion penetration probability might be large enough to keep a quasi-equilibrium accretion process, and the accretion strange star could be an X-ray pulsar (Bhattacharya et el. 1991, Nagase 1989). If the magnetic field is less strong ($B_p \sim 10^8$ gauss), and in case of low accretion rate (Low Massive X-ray Binaries), those two polar accreted crusts could be large enough to form a united crust, where the accretion process is mild, and the electric gap could prevent strong interactions between the crust and the strange matter. However, as the accreted matter piled up, the crust could be enough hot and dense to trigger the thermonuclear flash. In this case, the strange star could act as an X-ray burster (Lewin et al. 1993), and a lot of ions might also be pushed to the strange matter through the Coulomb barrier.

\section{A comparison between the magnetospheres}

If the magnetosphere of a bare strange star could be formed, a question is raised: what is the difference between the magnetospheres of bare strange stars and neutron stars? Our calculations show that, just outside of the surface (beyond $\sim 10^{-8}$ cm), the induced electric field can be large enough to control the magnetosphere. This situation is very similar to that for the neutron stars. Hence, Goldreich and Julian model (Goldreich \& Julian 1969) can also describe the magnetosphere of strange stars. There are two differences between the magnetospheres' properties of strange stars and of neutron stars. 
\begin{itemize}
\item Only electrons and positrons (no ions) are charged particles in the magnetosphere of a strange star. For the magnetosphere of a neutron star, the iron ions would be exist because the binding energy of a neutron star surface is too low to stop the irons flow out. 
\item In the magnetosphere of strange stars the RS inner gap (Ruderman \& Sutherlad 1975) can be formed easily. While in case of neutron stars, the RS inner gap is difficult to form, and the free-flow models (see e.g. Arons, 1983; Harding \& Muslimov 1998) will work.
\end{itemize}

Polar gap, as well as gap sparking, should be on the scene in the strange-star model for pulsars. Gil \& Cheng (1998) have noted the importance of polar gap sparking near the pulsars' surface. The short time scale sparking could be essential for the observed micro-pulses as well as (drifting) sub-pulses.

\section{The emission of strange star with a magnetosphere}

A wealth of observations has been collected for pulsars since the discovery of pulsars thirty years ago. Some important gaps still remain in our understanding of the emission process. General agreement begins and ends the statement that the very strong magnetic fields expected for neutron stars must play a prominent role (Sutherland 1979). Goldreich and Julian (1969) proposed a model to show that a rotating magnetic neutron star is surrounded by a charge-separated magnetosphere. Sturrock (1971) was the first to develop a comprehensive model for pulsar radiation, which suggests that an acceleration, immediately above the polar cap, will take place due to space-charge limited flow. Ruderman and Sutherland (1975) proposed an `inner gap' model. The model starts with the assumption that the binding of ions within the neutron star surface restrict the out free-flow of ions and a so-called `Inner gap' will be formed. In the inner gap ${\bf E}\cdot {\bf B}\neq 0$, $e^\pm$ particles can be produced through $\gamma-$B process, and can be accelerated relativisticly to produce $\gamma$-rays through curvature radiation (CR). Eventually, a cascade of pair production results in a discharge of the gap. Besides RS model, there are some models such as `slot gap' model (see e.g. Arans 1983), Beskin, Gurevich and Istomin (1988) model. But the `user friendly' nature of RS model is a virtue not shared by others (Shukre 1992). It is not uncommon to read contemporary observational papers in which the sole theoretical reference is to RS (1975!) (Michel 1992). Unfortunately, the RS model still faces great difficulties. From the theoretical point of view, some calculations (Hillebrandt and Muller 1976; Neuhauser et al. 1986, 1987; Kossl et al. 1988) show that the ion binding energy is at least one order less than what is required in the RS model, which means that the inner gap can not be formed. This is the so-called `binding energy problem'.

In another hand from observational point of view, Rankin's phenomenological work provides a much firmer basis for emission model than earlier models with only a hollow cone (Taylor and Stinebring 1986). The emission beams of radio pulsars can be divided into two (core and conal) emission components (Lyne and Manchester 1988) or three (core, inner conal and outer conal) emission components (Rankin 1983,1990,1993). But in RS model only one hollow cone emission component can be obtained. This is to say that there is serious conflict between the theory of RS model and the observations. Under an assumption of the inner gap sparking Qiao and Lin (1998) proposed an inverse Compton scattering model for radio emission of pulsars. In the model all of the core, inner cone and the outer cone emission components can be obtained. So most important problem faced by inner gap model is the `binding energy problem'. Now the `bare' strange star is not bare in fact, a magnetosphere can be formed due to the very strong electric field at the strange star surface, and the `binding energy problem' will be easy retrievable. Hence, the user-friendly nature of RS model can be survived.

For RS inner gap model, it is assumed that the magnetospheric charge density above the polar cap is positive (in the Goldreich-Julian model this means that the rotational angle velocity ${\bf \Omega}$ and the magnetic momentum ${\bf \mu}$ are anti-parallel). For neutron stars at which  ${\bf \Omega}$ and  ${\bf \mu}$ are parallel, the inner gap can not be formed and the inner gap model does not work. If the angle between $\Omega$ and $\mu$ of the neutron stars is distributed uniformly, the virtue above means that the RS inner gap model does not apply to half of the neutron stars. This is a strange virtue. If a magnetosphere can be formed around the `bare' strange star, this limitation does not exist again. Let's go to some details as follows for ${\bf \Omega}\cdot{\bf B}>0$.

For neutron stars, all of the electrons of irons in the crust are free as the electron Fermi energy is so high that the Electron Sea is only slightly perturbed by the nuclei Coulomb fields. The thickness $h$ of the crust for typical neutron star models is about several hundred meters, and the density of the crust $\rho = 10^6$g cm$^{-3}$ (Smith 1977). The number density of electron in the crust would be
$$
\rho_n = {26\rho \over 55.85\times 1.66\times 10^{-24}}.
$$
As electrons flow out from the polar cap surface, the radius of which is $R_{gap}$,
$$
R_{gap} = R \sqrt{2\pi R\over c P},
$$
where $R$ and $P$ are the radius and period of a neutron star, the matter below the cap should be positively charged, and the electrons out of this region will drift in across magnetic field lines at a velocity of $v_D$ (Jackson 1975),
$$
v_D = c {{\bf E}\times{\bf B}\over B^2},
$$
where $E$ is the electric field due to positive charging up, $B$ is the magnetic field. To consider the equilibrium case above, assuming the $\rho_x$ is the positive charged number-density, we come to
$$
\begin{array}{lll}
\eta_{NS}\equiv {\rho_x \over \rho_n} & = & {n_{GJ}\cdot B\over 4 \pi h \rho_n^2 e}\\
& \sim & 1.47\times 10^{-32},
\end{array}
$$
which means only a very slight departure of electron number density from $\rho_n$ can support a equilibrium free flow, here $n_{GJ}={\rho_{GJ}\over e}$, $e=4.8\times 10^{-10}$e.u.s is the electron electricity. So, an equilibrium of electron flows could be established because $\eta_{NS}$ is very small. In fact, there is a space charge separation in the neutron star, while, this effect is negligible in the above estimates.

Similarly, we consider the case for a bare strange star. From Fig. 4 (left), only electron above $z=z_c\sim 6\times 10^{-8}$cm could be flow out by induced electric field. The out flow rate $F_{out}$ could be
$$
\begin{array}{lll}
F_{out} & = & n_{GJ} \cdot \pi R_{gap}^2 \cdot c\\
& \sim & 1.4\times 10^{30} \;\; {\rm s^{-1}}.
\end{array}
$$
While, if the total electrons being available to flow are pulled out, the maximum discharging rate $F_{in}$ should be
$$
\begin{array}{lll}
F_{in} & = & c \int_{z_c}^{+\infty} n_e \cdot 2\pi R_{gap}\cdot dz\\
& \sim & 2.1\times 10^{32} \;\; {\rm s^{-1}}.
\end{array}
$$
In fact, a discharging rate in a real case should be much smaller than $F_{in}$ because the total electrons being available to flow are not pulled out at all. So, $ F_{in}$ and $ F_{out}$ are comparable, which result in the instability of the polar cap electron flow, and the inner gap similar to that of RS model could be possible.

The total numbers of electrons available to flow out from neutron star and from strange star are
$$
\begin{array}{lllll}
Q_{ns} & = & \rho_n h S_{gap} & = & 1.85\times 10^{42} h_4 \rho_6 R_6^3 P_1^{-1},\\
Q_{ss} & = & S_{gap} \int_{z_c}^{+\infty} n_e dz & = & 2.28\times10^{25} R_6^3 P_1^{-1},
\end{array}
$$
respectively, here $S_{gap}=\pi R_{gap}^2$. If the charge density of relativisticly flowing electrons at the surface is $\rho_{GJ}\mid_{r=R}$, then the times scales to pull all of this electrons out from neutron star and from strange star are
$$
\begin{array}{lll}
t_{ns} & = & 7.86 \times 10^9 h_4 \rho_6 B_{12}^{-1} P_1 \;\; {\rm seconds},\\
t_{ss} & = & 3.66\times10^{-5} B_{12}^{-1} P_1 \;\; {\rm seconds}.
\end{array}
$$
Therefore, the instability processes in the strange stars are more acute than that in the neutron stars, which could result in the depletions of charge above the polar caps.

\section{Conclusion \& Discussion}

It is shown that there might be a magnetosphere surrounding a bare strange star or namely isolated strange star. `Bare' strange stars might not be bare at all, which could be observed as rotation-powered pulsars. It is difficult to form an accretion crust around an isolated strange star. Strange stars with accretion crusts could be formed in binary systems, which can act as X-ray pulsars and X-ray bursters. 

There are some advantages if the strange stars (rather than neutron stars) are employed to work for the pulsars' radio emission:

1. It is easy to understand why the K-shell lines of iron have never be observed in X-ray emission bands of radio pulsars. The iron emission lines at $\sim 6.4$ keV and absorption edges at $\sim 7.3$ keV have been observed in many accretion-powered X-ray pulsars (Nagase 1989). Absorption lines at 5.7 keV or 4.1 keV have also been detected in several X-ray bursters (Lewin et al.1993), which could be considered as the iron element origin. While, there is not any observational signature of iron lines in the X-ray emission of rotation-powered pulsars (Cheng et al. 1998, Becker \& Trumper 1997, Thompson 1996), although the $e^\pm$ annihilation line in Carb have been observed and well explained (Zhu \& Ruderman 1998; Agrinier et al. 1990; Massaro et al. 1991). If pulsars are rotating magnetized neutron stars as universally believed, the binding energy per ion in the neutron star surface is too low to support the inner gap scenario, hence the ions will free-flow from the surface (Neuhauser et al. 1986,1987; Harding and Muslimov 1998). Therefore the composition of iron in the magnetosphere should be un-negligible, and it is hard to explain why the emission line has never been observed. But for strange star, the open field line regions of a strange star magnetosphere consists of $e^\pm$ pairs only, no iron can result in the radative processes. So, one of the discrimination criteria for strange star and neutron star in observation is to seek the iron lines of rotation-powered pulsars in X-ray bands.

2. It is easy to resolve the so-called `binding energy problem'. The RS model(Ruderman \& Sutherland 1975) has come closest to enabling comparison of observations and theory ( Radhakrishnan 1992). A fatal point of the model is that there is an inner gap above the polar cap, which inquires the binding energy of the positive ions large enough (larger than 10 keV) to restrict them from free-flowing out. However, the iron bounding energy can not be large enough to support the RS gap of neutron star. The calculations by Hillebrandt and Muller (1976), Muller (1984), Jones (1985,1986), Neubauser et al.(1986,1987), Kossl et al.(1988) show that the ion binding energy at least one order less than what is required in the RS model. The irons have its lowest energy state as unbound atoms rather than the chains. This is so-called `binding energy problem', which is faced if the signal of pulsars is coming from neutron stars. In case of strange star, as the positive charged quark matter near the surface is held by strong interaction, the binding energy should be approximately infinity when ${\bf \Omega}\cdot {\bf B} < 0$. Because of the attraction of the quark matter and the quasi-equilibrium of electric current, the electrons could not be easily pulled out when ${\bf \Omega}\cdot {\bf B} > 0$.

3. The strange star model for pulsars could be employed for ${\bf \Omega}\cdot {\bf B} < 0$ as well as ${\bf \Omega}\cdot {\bf B} > 0$. Is there any striking difference in observation and theory between these two situations of ${\bf \Omega}\cdot {\bf B} < 0$ and ${\bf \Omega}\cdot {\bf B} > 0$? This is an open question.

What's more, it is easier to collapse during the last stage of the supernova explosion if a strange quark matter star, rather than a neutron star, leaves over. During the last stages of the collapse of a supernova core, a shock wave will form and move outward due to the very stiff nuclear equation of states when the central density exceeds the nuclear-matter density. As the collapse continues, the phase transitions from nuclear matter to two-flavor quark matter and from two-flavor quark matter to three-quark matter may occur (Dai et al 1995). After the conversions, the neutrino energy in the whole collapse core increases obviously, which could result in the enhancement of both the probability of success for supernova explosion and the energy of the revived shock wave.

A critical point to distinguish neutron star and strange star is that the strange star have very high Coulomb barrier which can support a large body of matter, while, the neutron stars do not have. As the bare strange stars acting as the radio or $\gamma-$ray pulsars are very similar to the neutron stars (the differences of rotation periods and cooling curves between them are hard to be found), we suggest to search the differences between strange star and neutron star in accretion binaries, especially for the bursting X ray pulsar GRO J1744-28 (Strickman et al. 1996). 

One of the most serious difficulties is the possibility of strange star glitch. An important element of the theory (Pines \& Alpar 1985) to explain the observed glitches by neutron star is the ability of superfluid neutrons to move freely across magnetic field lines, while, there seems no neutral particle in a strange star. Nevertheless, perhaps, a strange star model on glitch might be developed by further research.

There might be no neutron stars in the galaxy (Alcock et al. 1986), all of the observed pulsars are the strange stars. If this is true, the strange matter hypothesis could be right, which can help us to interpret the properties of the six flavor quarks and the composition of the universe.

\begin{center}
{\Large \bf APPENDIX The electric state of a strange star}
\end{center}

For a static and nonmagnetized strange star, the properties of strange quark matter are determined by the thermodynamic potentials $\Omega_i$ (i = u, d, s, e) which are functions of chemical potential $\mu_i$ as well as the strange quark mass, $m_s$, and the strong interaction coupling constant $\alpha_c$ (Alcock et al. 1986). By assuming weak interaction chemical equilibrium and overall charge neutrality, we come to
$$
\begin{array}{lll}
\mu_d & = & \mu_s = \mu,\\
\mu_e & + & \mu_u = \mu,\\
n_e & = & (2n_u - n_d - n_s)/3,\\
n_i & = & -{\partial \Omega_i \over \partial \mu_i},
\end{array}	\eqno(A1a)
$$
and the total energy density $\rho$ reads
$$
\rho = \sum_{i=1}^4 (\Omega_i+\mu_i n_i)+B,	\eqno{A1b}
$$
where $B$ is the bag constant, and $\Omega_i$ referred to the Appendix in the paper by Alock et al.(1986). The above equations (A1a-A1b) have only one free independent parameter, $\mu$, and establish the relations between \{$\rho, \mu_i, n_i$; i = 1,2,3,4 for u, d, s, e, respectively\} (9 equations for 9 quantities).

\begin{figure}
$$\psfig{figure=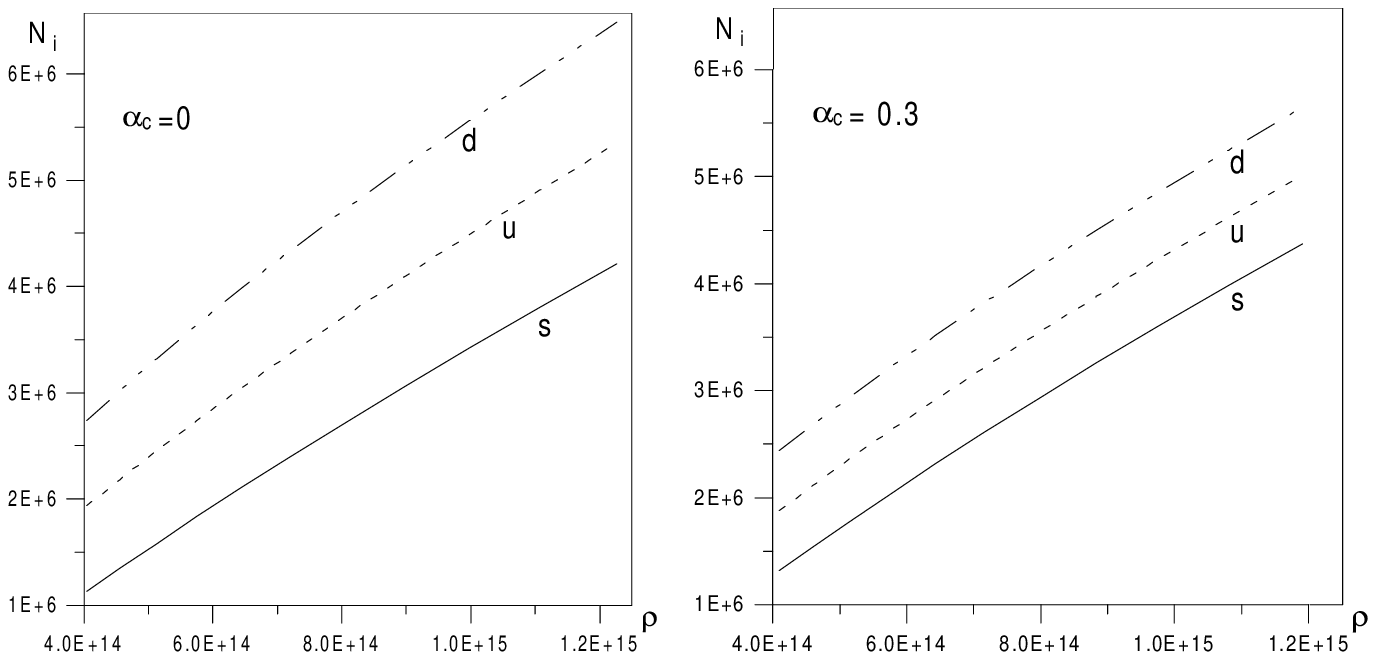,width=8cm,height=4cm,angle=0}$$
\caption[]{
 The number densities of u, d, and s quarks, $N_u$, $N_d$, $N_s$, are functions of total energy density $\rho$ (in g/cm3). $N_i$ refers to one of $N_u$, $N_d$, and $N_s$, which are in unit of particle number per cm$^3$. The couple constant $\alpha_c$ is chosen to be 0 (left) and 0.3 (right), respectively.
\label{Fig.A1}}
\end{figure}

\begin{figure}[h]
$$\psfig{figure=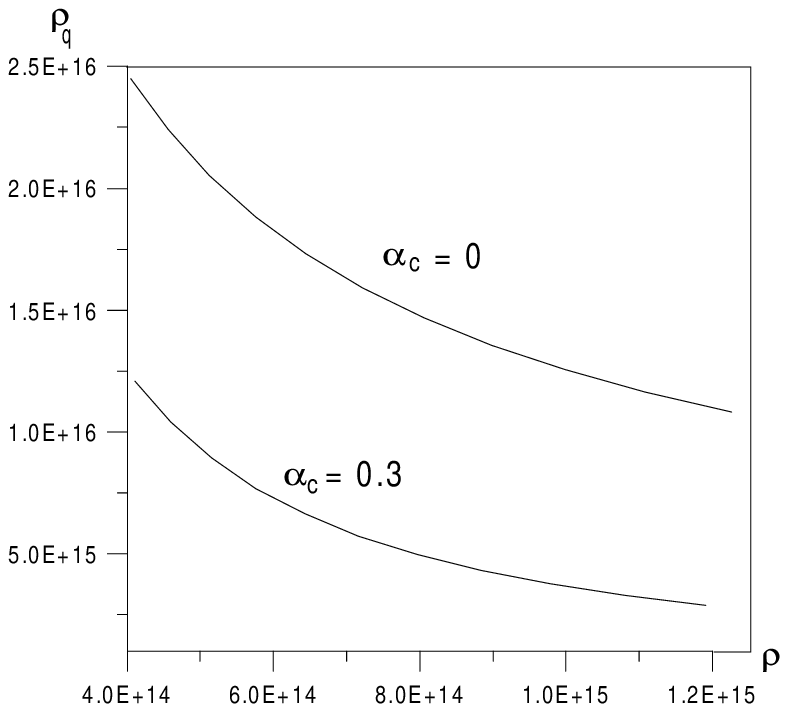,width=5cm,height=5cm,angle=0}$$
\caption[]{
The quark charge density $\rho_q$ (in unit of Coulomb per cm$^3$) decreases as a function of the total energy density $\rho$. The coupling constant $\alpha_c$ is chosen to be 0 and 0.3, respectively.
\label{Fig.A2}}
\end{figure}

The calculation results from equ.(A1) are shown in Fig.2, and Fig.3, where the number densities of u, d, s quarks, and the quark charge density $\rho_q$ (in unit of Coulomb per cm$^3$) are varied as a function of total energy density $\rho$. In the computation, we choose $B = (145 MeV)^4$, $m_s = 200$ MeV, and the renormalization point $\rho_R = 313$ MeV, both for $\alpha_c = 0$ and $\alpha_c = 0.3$. As $\rho$ has a mild rise variation from the outer part to the interior of a strange star (Alcock et al. 1986), the number density of u, d, and s quarks increase almost in a same degree. However, the equilibrium quark charge density $\rho_q$ changes significantly, as $\rho$ increases (Fig.3), which means the number of equilibrium electrons becomes smaller as one goes to a deeper region of a strange star. For a strange star with a typical pulsar mass 1.4 M$_\odot$, the total energy $\rho$ has a very modest variation with radial distance of strange star (Alcock et al 1986), from $\sim4\times10^{14}$g cm$^{-3}$ (near surface) to $\sim 8\times 10^{14}$g cm$^{-3}$ (near center), therefore the quark charge density $\rho_q$ would be order of $10^{15}$ ($\alpha_c = 0.3$) to $10^{16}$ ($\alpha_c = 0$) Coulomb cm$^{-3}$. Physically, as the Fermi energy of quarks becomes higher (for lager $\rho$), the effect due to $m_s \not= 0$ would be less important, hence, the charge density should be smaller.

Since the quark matter are bounded up through strong interaction (the thickness of the quark surface will be of order 1 fm), and the electrons are held to the quark matter electrically, hence the electrons distribution would extend beyond the quark matter surface. A simple Thomas-Fermi model has been employed to solve for this distribution (Alcock el al. 1986), and the local charge distribution can be obtained by Poisson's equation
$$
{d^2 V\over dz^2} = 
\left\{    \begin{array}{ll}
{4\alpha\over 3\pi}(V^3-V_q^3) & z\leq 0,\\
{4\alpha\over 3\pi} V^3 & z > 0,
\end{array}     \right.  \eqno(A2)
$$
where z is a measured height above the quark surface, $\alpha$ is the fine-structure constant, $V_q^3/(3\pi^2)$ is the quark charge density, $V/e$ is the electrostatic potential, and the number density of electrons is given by
$$
n_e = {V^3\over 3\pi^2}. \eqno(A3)
$$
Physically, the boundary condition for equ.(A2) are
$$
\begin{array}{ll}
z \rightarrow -\infty: & V \rightarrow V_q, dV/dz \rightarrow 0;\\
z \rightarrow +\infty: & V \rightarrow 0,   dV/dz \rightarrow 0.
\end{array}
$$
By a straightforward integration of equ.(A2), we get
$$
{dV \over dz} = 
\left\{    \begin{array}{l}
-\sqrt{2\alpha\over 3\pi} \cdot \sqrt{V^4-4V_q^3V+V_q^4},\\
-\sqrt{2\alpha\over 3\pi} \cdot V^2.
\end{array}     \right.  \eqno(A4)
$$
From equ.(A4), the continuity of $dV/dz$ at $z = 0$ requires $V(z=0) = 3V_q/4$(Alcock et al. 1986), and we can consider the solution of equ.(A4) for $z > 0$ by
$$
{dV \over dz} = -\sqrt{2\alpha\over 3\pi} V^2,  {\rm boundary}: V(z=0) = {3\over 4}V_q,	\eqno(A5)
$$
hence,
$$
V={3V_q\over \sqrt{6\alpha\over\pi}V_qz+4}.\;\;({\rm for} z > 0) \eqno(A6)
$$

It is interesting that, although the electric field near the surface is order of $10^{17}$ V cm$^{-1}$, the electric field decrease very quickly above the quark surface. We choose $V_q = 20$ MeV (hence $\rho_q = 270.19$ MeV$^3 = 5.63\times10^{15}$ Coulomb cm$^{-3}$), and the calculated electric field as a function of $z$ is shown in figure 4 (left), where the electric field is $4.5\times 10^{17}$ V cm$^{-1}$ when $z = 0$. Also the electron charge density curve is drawn in Fig.4 (right), which decreases from ${3\over 4}V_q = 4.2\times 10^{15}$ Coulomb cm$^{-3}$ at the surface to $3.3\times 10^{-9}$ Coulomb cm$^{-3}$ when $z = 3\times 10^{-3}$ cm.

\begin{figure}
$$\psfig{figure=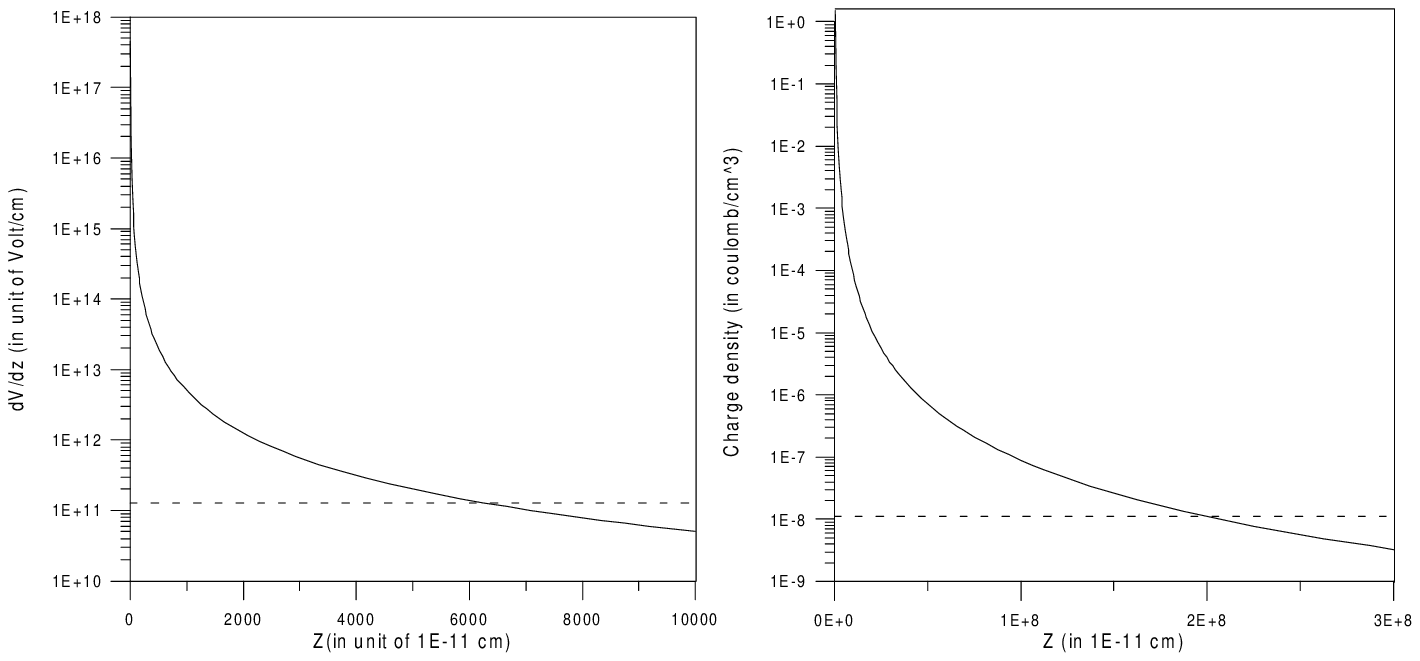,width=8cm,height=4cm,angle=0}$$
\caption[]{
The electric field and the electron charge density variation curves as function of $z$ (a space coordinate measuring height above the quark surface). The
 dashed lines are for the unipolar induced electric field and charge separated density.
\label{Fig.A3}}
\end{figure}

\bigskip
\noindent{\bf Acknowledgments:}
We are very grateful to Prof. T. Lu and Q. H. Peng for their valuable discussion and encouragement. We would like to thank B.Zhang, B.H.Hong and J.F.Liu for helpful discussions. This work is partly supported by NSF of China, the Climbing Project-National Key Project for Fundamental Research of China, the Doctoral Program Foundation of Institution of Higher Education in China, and the youth scientific foundation of Peking University.

\noindent{\bf References:}
\begin{description}

\item[]\hspace{-3mm}
Agrinier,B., et al. 1990, ApJ, 355,645
\vspace{-3mm}
\item[]\hspace{-3mm} 
Alcock,C., Farhi,E., \& Olinto,A. 1986, ApJ, 310, 261
\vspace{-3mm}
\item[]\hspace{-3mm} 
Arons,J. 1983, ApJ, 276, 215
\vspace{-3mm}
\item[]\hspace{-3mm} 
Becker,W., \& Trumper,J. 1997, A\&A, 326, 682
\vspace{-3mm}
\item[]\hspace{-3mm} 
Benvenuto,O.G., Lugones,G. 1995, Phys. Rev., D51, 1989
\vspace{-3mm}
\item[]\hspace{-3mm} 
Benvenuto,O.G., \& Vucetich,H. 1991, Nucl. Phys., (Proc. Suppl.) B24, 160
\vspace{-3mm}
\item[]\hspace{-3mm} 
Beskin,V.S., Gurevich,A.V., \& Istomin,Y.N. 1986, ApSS, 146, 205
\vspace{-3mm}
\item[]\hspace{-3mm}
Bhattacharya,D., van den Heuvel,E.P.J. 1991, Phys.Rep., 203, 1
\vspace{-3mm}
\item[]\hspace{-3mm} 
Broderick, J.J., et al. 1998, ApJ, 192, L71
\vspace{-3mm}
\item[]\hspace{-3mm} 
Bodmer,A.R. 1971, Phys. Rev., D4, 1601
\vspace{-3mm}
\item[]\hspace{-3mm} 
Cheng,K.S., Ho,C., \& Ruderman,M. 1984, ApJ, 300, 500
\vspace{-3mm}
\item[]\hspace{-3mm}
Cheng,K.S., Gil,J., \& Zhang,L. 1998, ApJ, 493, L35
\vspace{-3mm}
\item[]\hspace{-3mm} 
Dai,Z.G., \& Lu,T. 1986, Zeit. F. Physik A, 355, 415
\vspace{-3mm}
\item[]\hspace{-3mm} 
Dai,Z.G., Peng,Q.H., \& Lu,T. 1995, ApJ, 440, 815
\vspace{-3mm}
\item[]\hspace{-3mm} 
Erber,T. 1966, Rev. Mod. Phys., 38, 626
\vspace{-3mm}
\item[]\hspace{-3mm} 
Farhi,E., \& Jaffe,R.L 1984, Phys. Rev., D30, 2379
\vspace{-3mm}
\item[]\hspace{-3mm} 
Fawley,M., Arons,J., \& Scharlemann,E.T 1977, ApJ, 217, 227
\vspace{-3mm}
\item[]\hspace{-3mm} 
Flower,E.G., et al. 1977, ApJ, 215, 291
\vspace{-3mm}
\item[]\hspace{-3mm} 
Frieman,J.A., \& Olinto,A. 1989, Nature, 341, 633
\vspace{-3mm}
\item[]\hspace{-3mm} 
Gil,J.A., \& Cheng,K.S. 1998, A\&A, submitted
\vspace{-3mm}
\item[]\hspace{-3mm} 
Goldreich,P. 1972, in The Physics of Pulsars, ed. A.M.Lenchek, Gordon and 
Breach, Science Publisher, 151
\vspace{-3mm}
\item[]\hspace{-3mm} 
Goldreich,P., \& Jullian,W.H. 1969, ApJ, 157, 869
\vspace{-3mm}
\item[]\hspace{-3mm} 
Harding,A.K., \& Muslimov,A.G. 1998, submitted to ApJ
\vspace{-3mm}
\item[]\hspace{-3mm} 
Hewish,A., Bell,S.J. et al. 1968, Nature, 217, 709
\vspace{-3mm}
\item[]\hspace{-3mm} 
Hillebrands,W., Muller,E. 1976, ApJ, 207, 589
\vspace{-3mm}
\item[]\hspace{-3mm} 
Jackson,J.D. 1975, {\it Classical Electrodynamics}, John Wiley \& Sons
\vspace{-3mm}
\item[]\hspace{-3mm}
Jones,P.B. 1985, Phys. Rev. Lett., 55, 1338
\vspace{-3mm}
\item[]\hspace{-3mm} 
Jones,P.B. 1986, M.N.R.A.S., 218, 477
\vspace{-3mm}
\item[]\hspace{-3mm} 
Kettner,Ch., Weber,F., \& Weigel,M.K. 1995, Phys. Rev., D51, 1440
\vspace{-3mm}
\item[]\hspace{-3mm} 
Kossl,D., Wolff,R.G., Muller,E., Hillebrands,W. 1988, A\&A, 205, 34
\vspace{-3mm}
\item[]\hspace{-3mm} 
Landau,L. 1932, Phys.Z.Sowjetunion, 1, 285,
\vspace{-3mm}
\item[]\hspace{-3mm} 
Lewin,W.H.G., van Paradijs,J., Taam,R.E. 1993, Space Sci. Rev., 62, 223
\vspace{-3mm}
\item[]\hspace{-3mm}
Lyne,A.G., \& Manchester,R.N. 1988, M.N.R.A.S., 134, 477
\vspace{-3mm}
\item[]\hspace{-3mm}
Massaro,E., et al. 1991, ApJ, 376, L11
\vspace{-3mm}
\item[]\hspace{-3mm}
McLerran,L. 1986, Rev. Mod. Phys., 58, 1021
\vspace{-3mm}
\item[]\hspace{-3mm} 
Melrose,D.B. 1995, J.Astrophy.Astr., 16, 138
\vspace{-3mm}
\item[]\hspace{-3mm} 
Michel,F.C. 1991, {\it Theory of Neutron Star Magnetospheres}, Univ. Chicago Press
\vspace{-3mm}
\item[]\hspace{-3mm} 
Michel,F.C. 1992, in The magnetospheric structure and emission mechanisms of radio pulsars, eds. Hankins,H., Rankin,M., \& Gil,A., Pedagogical Univ. press, 405
\vspace{-3mm}
\item[]\hspace{-3mm}
Muller,B. 1995, Rep. Prog. Phys., 58, 611
\vspace{-3mm}
\item[]\hspace{-3mm}
Muller,E. 1984, A\&A, 130, 415
\vspace{-3mm}
\item[]\hspace{-3mm} 
Nagase,F., 1989, Publ. Astron. Sco. Japan, 41, 1
\vspace{-3mm}
\item[]\hspace{-3mm} 
Neuhauser,D., Langanke,K., \& Koonin,S.E. 1986, Phys. Rev., A33, 2084
\vspace{-3mm}
\item[]\hspace{-3mm} 
Neuhauser,D., Koonin,S.E., \& Langanke K. 1987, Phys. Rev., A36,4163
\vspace{-3mm}
\item[]\hspace{-3mm} 
Oppenheimer,J.R., \& Volkoff,G.M. 1939, Phys. Rev., 55, 374
\vspace{-3mm}
\item[]\hspace{-3mm} 
Pines,D., \& Alpar,M.A. 1985, Nature, 316, 27
\vspace{-3mm}
\item[]\hspace{-3mm} 
Radhakrishnan,V. 1992, in The magnetospheric structure and emission 
mechanisms of radio pulsars, eds. Hankins,H., Rankin, M., and Gil, A., 
Pedagogical Univ. press, 367
\vspace{-3mm}
\item[]\hspace{-3mm}
Qiao,G.J., \& Lin,W.P. 1998, A\&A, 333, 172
\vspace{-3mm}
\item[]\hspace{-3mm}
Rankin,J.M. 1983, ApJ, 274, 333
\vspace{-3mm}
\item[]\hspace{-3mm}
Rankin,J.M. 1990, ApJ, 352, 247
\vspace{-3mm}
\item[]\hspace{-3mm}
Rankin,J.M. 1993, ApJ, 405, 285
\vspace{-3mm}
\item[]\hspace{-3mm} 
Ruderman,M.A., \& Sutherland,P.G. 1975, ApJ, 196, 51
\vspace{-3mm}
\item[]\hspace{-3mm} 
Schaffner,J. et al. 1993, Phys. Rev. Lett., 71, 1328
\vspace{-3mm}
\item[]\hspace{-3mm} 
Shibata,S. 1991, ApJ, 378, 239
\vspace{-3mm}
\item[]\hspace{-3mm} 
Shukre,C.S. 1992, in The magnetospheric structure and emission mechanisms of radio pulsars, eds. Hankins,H., Rankin,M., \& Gil,A. Pedagogical Univ. press, 412
\vspace{-3mm}
\item[]\hspace{-3mm}
Smith,F.G. {\it Pulsars}, 1977, Cambridge Univ. Press
\vspace{-3mm}
\item[]\hspace{-3mm} 
Strickman,M.S. et al. 1996, ApJ, 464, L131
\vspace{-3mm}
\item[]\hspace{-3mm} 
Sturock,P.A. 1971, ApJ, 164, 529
%\vspace{-3mm}
%\item[]\hspace{-3mm} 
%Sutherland,P.G. 1979, %------------------------------------------------------------------------->>%> ???
\vspace{-3mm}
\item[]\hspace{-3mm} 
Thompson,D.J. 1996, in Pulsars: Problem \& Progress, ed. S.Johnston et al., 307
\vspace{-3mm}
\item[]\hspace{-3mm}
Taylor,J.R., \& Stinebring,D.R. 1986, Ann. Rev. Astr. Astrophys. 24, 285
\vspace{-3mm}
\item[]\hspace{-3mm} 
Wang,F.Y.-H., \& Halpern,J.P. 1997, ApJ, 482, L159
\vspace{-3mm}
\item[]\hspace{-3mm} 
Wang,Q.D., \& Lu,T. 1984, Phys. Lett., B148, 211
\vspace{-3mm}
\item[]\hspace{-3mm} 
Witten,E. 1984, Phy. Rev., D30, 272
\vspace{-3mm}
\item[]\hspace{-3mm} 
Zhu,T., \& Ruderman,M. 1998, ApJ, 478, 701

\end{description}

\end{document}